\documentclass[aip,pof,numerical,reprint,showpacs]{revtex4-1}

\usepackage{amsmath,amssymb}
\usepackage{graphicx}
\usepackage{color}
\usepackage{mdwlist}

\begin{document}  
   
\title{When sound slows down bubbles}
\author{R\'emi Dangla}
\author{C\'edric Poulain}
\email{cedric.poulain@cea.fr}
 \affiliation{ CEA, DEN, F38054, Grenoble, France.}

\date{\today}

\begin{abstract}

We report experimental observations that a bubble rising in water in the presence of a sound field is significantly slowed down, even at moderate acoustic pressures.

  We measure the mean rise velocity of bubbles under various acoustic forcings and show this effect occurs if the noise spectrum matches or overlaps bubble resonance. We render surface oscillations and translational movements of bubbles using high speed video imaging and thereby identify Faraday waves on the bubble wall as the cause for the velocity reduction. The associated mechanisms are discussed in terms of induced forces.

\end{abstract}
 
  \maketitle 
  

The velocity at which buoyant gas bubbles rise to the surface of a liquid is key to geophysical and industrial processes~\cite{Mudde05}. 
Hence, considerable effort has been devoted to understanding the hydrodynamics of a single bubble~\cite{Magnaudet00}, underlying the major role played by the shape and the wake on its buoyant rise.

Bubbles are also known to be highly sound-sensitive objects~\cite{Leighton_book}, a property which also has implications in various fields: from sound propagation in oceans~\cite{Clay77}, to sonoluminescence and sonochemistry~\cite{Crum_book}, or even for biomedical applications~\cite{Marmottant03}.  Acoustically, a bubble responds to acoustic pressure fluctuations as a peculiar resonator since its resonant Minnaert frequency $f_M$ (Hz) obeys 
\begin{equation} 
f_M \approx 3.26/ R \mbox{ , valid for bubble radius $R > 10^{-6}$~m,  }
\label{eq:minnaert}
\end{equation}
and corresponds to a forcing wavelength in water much greater than the bubble size, such that the bubble undergoes radial oscillations. Yet, beyond a critical acceleration of the bubble wall, the parametric Faraday instability is triggered and shape oscillations (surface waves) develop~\cite{Faraday31,Ramble98}. Furthermore, because a bubble is a sharp oscillator, this threshold is lowest in terms of forcing pressure if the bubble is excited near resonance. Acoustic forcing of these waves has interesting applications such as bubble sizing~\cite{Phelps97}, electrodeposition patterning~\cite{Offin06} or mass transfer enhancements~\cite{Birkin01}.

Although most of the above cited applications involve buoyant bubbles rising in noisy environments, fewer studies~\cite{Eller70,Larraza00,Rensen2001} concern these coupled aspects. The shape of a bubble being key to its hydrodynamics, surface waves are expected to significantly modify the bubble rise. However, existing results focus on the influence of high intensity sound fields on bubble motion, in which case effects of the Faraday instability, if triggered, overlap with other mechanisms such as the acoustic radiation force. 

Hence, it is logical to address the following question :  ``what is the effect of a \emph{low intensity, permanent and resonant} sound field on the rise of a bubble?''
 
In this letter, we report experiments studying the interaction of a single rising bubble with a resonant sound field. Rise velocities with and without sound are compared by time of flight measurements to capture effects of the acoustics. Our results are discussed to identify the physics at play but the complete theoretical modeling is beyond the scope of this letter.


\begin{figure}[!htb]
\includegraphics[height=10cm]{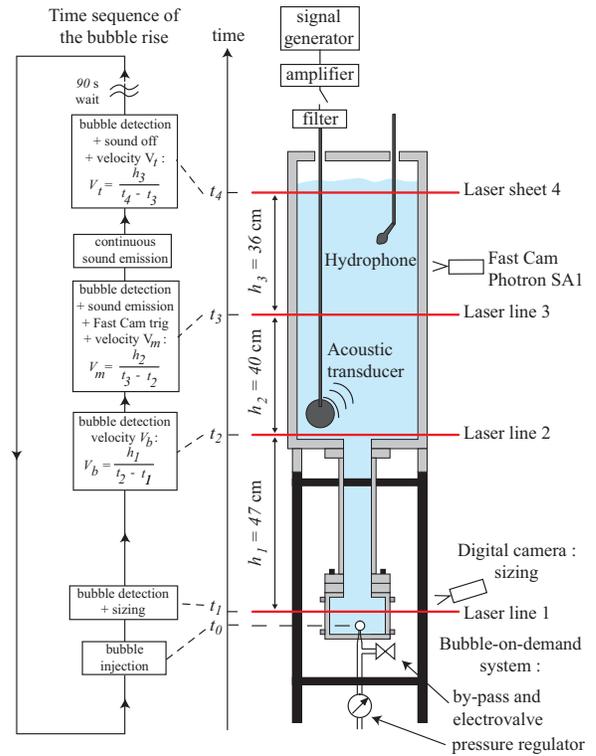}
\caption{\textbf{Right} : Sketch of the experimental set-up featuring the bubble-on-demand system, the 4 lasers that detect bubble crossings to measure rise velocities and to trigger events, the sound emission chain and the digital cameras. \textbf{Left} : the time sequence of a bubble rise is decomposed in key steps including injection, sizing, velocity measurements and sound emission triggering.}
\label{fig:expSetup}
\end{figure}

The experimental set-up is sketched on Fig.~\ref{fig:expSetup}. Air bubbles are produced at the bottom of a tall Plexiglas water tank: a cubic visualization cell ($9$~cm wide)  linked to a round tube ($6$~cm  i.d, $40$~cm height) leads to a squared vessel ($30 \times 30 \times 90$~cm). Experiments are carried out using purified deionized water at room temperature ($T\approx25^0$~C). The vessel is sealed to limit contamination.  Single bubbles are released from a capillary silica tube ($20$~$\mu$m i.d., $90$~$\mu$m o.d.) connected to a fixed pressure source.  A home made system (by-pass and electrovalve) enables on-demand bubble production every $90$~s to ensure that the previous bubble and its wake have vanished. Upon release, the bubble is detected and sized at the bottom of the tank by  a digital camera synchronized to a stroboscopic flash. The bottom bubble radius $R_b$ is extracted by live image processing using a MatLab routine ($1\%$ accuracy). Very good reproducibility is obtained since the diameters of successive bubbles are identical within the  measurements error margin.
 
During rise, 4 lasers (3 lines and 1 sheet) placed at heights $5$~cm, $53$~cm, $93$~cm and $129$~cm above the injection nozzle detect bubble crossings, giving 4 trigger times and thus 3 rise velocity measurements by time of flight ($V_b$, $V_m$ and $V_t$ for bottom, middle and top respectively - see Fig.~\ref{fig:expSetup} for details). Crossing detections are also used as triggers : for sound emission and high speed camera acquisition by the laser 3 and for sound deactivation by the top laser sheet.

Sound is produced by an ITC 1001 spherical transducer (an omnidirectional projector) powered by an audio amplifier (HPA A2400) followed by a transformer (ratio $ 4.5$). Monochromatic or band filtered white noise signals can be generated using a B\&K  3310 Pulse system. The projector is placed at the bottom of the top vessel in most experiments. Sound pressure signals  are recorded by a hydrophone (B\&K type 8103) connected to a charge amplifier ( B\&K Nexus). The tank being a resonant cavity, the sound field along the vertical centerline of the tank is mapped prior to each series of experiments involving monochromatic sound. The hydrophone is then placed near a wave crest $2$~cm away from the centerline to give a measure of the sound pressure in the resonant tank throughout the experiment~\footnote{Sound pressures are given in RMS values to enable a direct comparison between monochromatic and band filtered noise values}. A high speed video camera (Photron SA1) captures the rise and oscillations of the bubble in synchronization with the hydrophone sound pressure signal using a dedicated synchronized fast-acquisition system (Photron MCDL).
    
The experimental protocol is the following: First, bubbles are produced until stability of the injection nozzle is reached, setting a reference bubble radius $R_b^0$ for the rest of the series. Assuming an isothermal hydrostatic expansion of the bubbles, the radius $R_t$ at the height of laser 3 is predicted and injected into a refined version of \eqref{eq:minnaert} (cf. \cite{Leighton_book} p.183.) to calculate the resonant frequency $f_M$, taking dissipative effects into account. Reverberation effects can be shown to be negligible in our configuration~\cite{Leighton02}. Afterwards, a series of bubbles are released without acoustic forcing to determine the three reference quiescent rise velocities ($V_b^0$, $V_m^0$ and $V_t^0$). Only then is sound emission activated and the effect of acoustics are probed. The bottom $V_b$ and middle $V_m$ velocities are still measured in a quiescent environment. Along with the bubble radius $R_b$, they are indicators of reproducibility as they should remain equal to the reference values $R_b^0$, $V_b^0$ and $V_m^0$, and are used to rule out random events such as changes in bubbling regime. Finally, the effect of acoustics on the bubble ascent is captured by the top rise velocity $V_t$, measured above the laser 3 that triggers sound emission.

\begin{figure}[!htb]

\includegraphics[width=8cm]{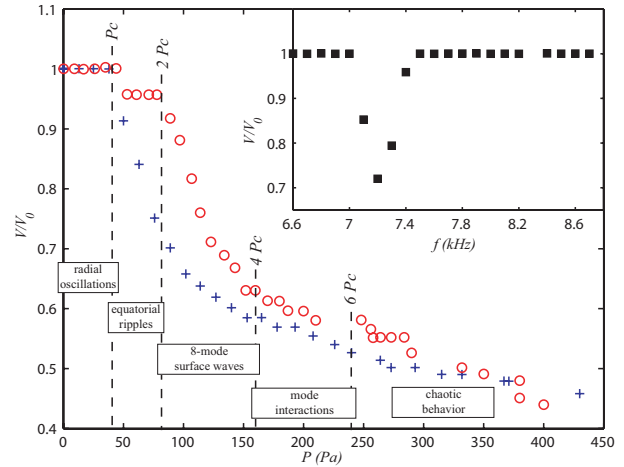}
\caption{Bubble velocity $V_t$ normalized by the quiescent reference velocity $V_t^0$, as a function of the root mean square (RMS) sound  pressure $P$ for monochromatic (+) and band filtered white noise (o) emission. For the monochromatic case (+), $R_b^0=452$~$\mu$m, $V_t^0=19.2$~cm$\cdot$s$^{-1}$, $f_M = 7.04$~kHz and $f=7.2$~kHz. For the filtered white noise, $R_b^0=446$~$\mu$m, $V_t^0=18.9$~cm$\cdot$s$^{-1}$, $f_M = 7.13$~kHz, central frequency $f=7.2 $~kHz with a span $\Delta f=1.6$~kHz. Pressure bands relate the different surface oscillation regimes to the velocity decrease.
\\Insert: $V_t/V_t^0$ as a function of the acoustic frequency $f$ for a monochromatic forcing: $R_b^0=447$~$\mu$m, $V_t^0=19.0$~cm$\cdot$s$^{-1}$, $f_M = 7.12$~kHz,  $P=114$~Pa.}
\label{fig:velocity_pressure}
\end{figure}

We run experiments in which bubbles are forced at a fixed frequency $f$ near resonance $f_M$ and at varying sound levels. The top rise velocity $V_t$ normalized by the reference value $V_t^0$ is plotted in Fig.~\ref{fig:velocity_pressure} as a function of the acoustic pressure $P$ for both monochromatic sound and band filtered white noise of fixed frequency span. The curves featured on Fig.~\ref{fig:velocity_pressure} are typical of all conducted experiments, in which bubble radii ranged from  $300$~$\mu$m to $500$~$\mu$m. Firstly, we observe that the acoustic forcing at resonance reduces rise velocity by up to 30\% for  moderate noise levels of order $P \approx 100$~Pa, and by more than 50\% for $P > 300$~Pa.  Secondly, there is always a clear pressure threshold $P_c$ upon which acoustic effects come into play, around $P = 50$~Pa for the plotted measurements. This threshold does not differ significantly from monochromatic to band filtered white-noise cases. Furthermore, we recover the shape of the curves whatever the projector's position, indicating that only the value for the threshold $P_c$ varies with the acoustic forcing characteristics (projector position and spectrum).

To test the frequency sensitivity of the phenomenon, we excite a series of bubbles by monochromatic forcings at a constant sound pressure $P$ above the threshold $P_c$ for velocity reduction and at various frequencies around resonance. Insert in Fig.~\ref{fig:velocity_pressure} shows that bubbles are only slowed down by acoustic forcings near resonance for a narrow bandwidth of $300$~Hz approximately, which is of the order of the bubble's resonant bandwidth $\Delta f_0 = 280$~Hz~\cite{Leighton_book}. We also verify that forcing at harmonics and subharmonics of the resonant frequency have no effect on the rise velocity. We obtain similar results using band-filtered white noise forcings but the added parameter of frequency span leads to a more complex parametric analysis. 

Hence, the velocity reduction induced by the acoustic forcing is shown to be a threshold effect restricted to a narrow frequency band around resonance. Other phenomenons involving acoustic forcing of bubbles share such onset properties, such as subharmonic emissions of resonant bubbles~\cite{Phelps97} and mass transfer enhancements~\cite{Birkin01}. In these problems, the origin for the threshold behavior is the parametric Faraday instability, which triggers surface waves on the bubble wall. Theoretical models and experimental measurements~\cite{Nabergoj79,Ramble98} predict the triggering of Faraday waves on bubbles of $500$~$\mu$m radius for sound pressures of approximately $P \approx 5 \cdot 10^{-2}/R = 100$~Pa, which is of the order of the threshold $P_c$ we measure for rise velocity reduction. To substantiate the link, we visualize in detail bubble behavior under acoustic forcing at resonance using high speed video imaging ($62500$~fps).

\begin{figure}[!htb]
\includegraphics[width = 9cm]{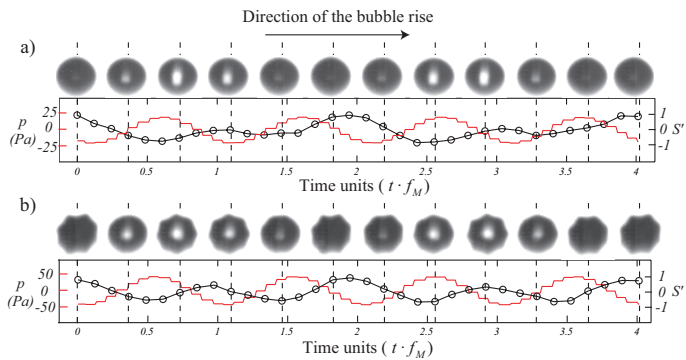}

\caption{Frame sequence for rising bubbles in (a) equatorial mode and (b) mode 8 oscillations. Curves below represent the instantaneous acoustic pressure $p(t)$ near the bubble ( - ) and the normalized projected area fluctuations $S'$ of the bubble (o). $R_b=420$~$\mu$m, $f=f_M=7.6$~kHz, (a) $P=P_c=15$~Pa and (b) $P=31$~Pa, $V_t^0=17.8$~cm$\cdot$s$^{-1}$, frame rate $62500$~fps.}
\label{fig:videosequence}
\end{figure}

In the case of monochromatic sound emission, the videos reveal a clear transition in bubble surface motion at the pressure threshold $P_c$. Below this onset value, bubbles oscillate radially while above, they undergo shape oscillations. The bubble geometric distortions are first localized near the equator, as shown in figure \ref{fig:videosequence}(a). Such equatorial modes have been observed for tethered bubbles and modeled in the framework of Faraday waves as self-focusing ripples~\cite{Maksimov08}. Here, this particular feature is visualized on rising bubbles.

As the acoustic pressure increases to a value of $P \approx 2\cdot P_c$, the pattern formed by the Faraday waves changes from equatorial distortions to $8$-mode oscillations, as shown on figure \ref{fig:videosequence}(b).  This observation is consistent with theoretical results\cite{Nabergoj79,Ramble98}, which predict the $8$-mode as the most unstable parametric surface wave for a non moving oscillating bubble of size $R=420$~$\mu$m.

Well above the pressure threshold, for $4<P/P_c<6$ approximately, multiple unstable modes interact and give rise to less regular geometric deformations and complex patterns. For even stronger forcing of typically $P/P_c>6$, the bubble oscillations are highly distorted, no longer axisymetric nor harmonic. In this regime, the rise velocity $V_t$ decreases to $50\%$ of the reference value $V_t^0$ and the bubble ``dances'': it follows an erratic path along the vertical and fragmentation occurs, as often observed in acoustic trapping experiments~\cite{Eller70}. Each regime is illustrated by a movie in the auxiliary material~\cite{Web_movies}.

In the video sequences under white noise emissions band-filtered around resonance, the bubble exhibits non-stationary volume and shape oscillations, as both the local frequency and amplitude of the sound wave that impinges on the bubble are time dependent. Therefore, the onset threshold for Faraday waves is only met transiently and the instability never reaches a steady state~\cite{Maksimov01}. Nonetheless, the surface oscillations of the bubbles can still be qualitatively described as sequences of equatorial ripples, n-mode oscillations and mode superpositions. 

Additional quantitative evidence is brought by the frequency of the surface oscillations which can be estimated from the video sequences and the time traces of the bubble's projected area and compared to the sound pressure signals, shown in Fig.~\ref{fig:videosequence}. It is clearly subharmonic of the forcing frequency, a key feature of the Faraday instability.

As a whole, these observations establish surface Faraday waves as the cause for the rise velocity reduction but the underlying mechanism needs to be discussed.
In our experiments, the bubble is sufficiently small that, in the absence of sound, it rises along a vertical line~\cite{Magnaudet00} at a velocity fixed by two competing forces: the buoyant driving force $\vec{F_b} = \rho \mathcal{V} \vec g $ in which $\rho$ is the density of water, $\mathcal V$ the bubble volume and $\vec g$ gravity, and a drag force $\vec{F_d} = -1/2 C_d \rho \pi R^2 \left| V \right| \vec V$ where $C_d$ is a shape-dependent drag coefficient~\footnote{We omit to detail the velocity dependence of the drag coefficient for simplicity} and $\vec V$ the bubble velocity.

When sound is produced below the Faraday threshold $P_c$, the bubble undergoes radial volume oscillations (a few percent in amplitude: $\delta R/R \propto 10^{-2}$), leading the force balance into a non-steady state. Hence, the rise velocity also oscillates around a mean value, such that the unsteady added mass~\cite{Ohl03} and history~\cite{Magnaudet98} forces come into play. Yet, we observe that the mean rise velocity is not altered. Indeed, the contribution of these forces to the mean momentum balance is at most $O(\delta R^2/R^2)$ when compared to the buoyancy force. Nevertheless, other effects have to be examined: \textit{First}, velocity fluctuations also couple with radial oscillations, giving rise to a streaming effect. Under the operating configuration, the streaming flow is of order~\cite{Longuet98} $1$~mm$\cdot$s$^{-1}$, which is negligible compared to the bubble rise velocity $V_t^0$. \textit{Second}, the sound field produces a radiation pressure on the bubble\cite{Doinikov98}. However, this force requires strong acoustics fields to be significant, as illustrated by the following scaling argument. Pressure fluctuations of amplitude $P'$ induce volume oscillations $V' \propto R/(\rho \xi f^2)\cdot P'$, where $\xi \propto 10^{-2}$ is the damping coefficient of the bubble and $f$ the forcing frequency~\cite{Leighton_book}. The pressure gradient scales as $\vec \nabla P \propto f/c \cdot P'$ with $c$ the sound velocity in water. At resonance, the forcing frequency $f$ is given by \eqref{eq:minnaert}, such that the radiation force relative scales as $F_r/F_b \propto <\mathcal V' \cdot \vec \nabla P > / F_b \propto 10^{-9} P^2/R $ relatively to buoyancy. Hence, the radiation force is significant for acoustic pressures above $P_r \approx 10^4/\sqrt{R}$ beyond the threshold for Faraday waves $P_c \approx 5 \cdot 10^{-2}/R $, provided bubbles are at least micron sized.

For sound pressures above $P_c$, the same forces compete but surface waves now superimpose onto the volume oscillations. The exact mechanism by which the Faraday waves reduce the rise velocity is difficult to identify given the number of forces acting upon the bubble and the complexity of the shape oscillations.  A likely candidate is the increase in the mean hydrodynamic drag $\vec{F_d}$ through modification of the drag coefficient $C_d$. 
Such variations are be significant: the drag coefficient of a cube can be twice that of an equivalent sphere, depending on its orientation. Here, given the observed angular shapes, $C_d$ always increases when the bubble departs its spherical geometry, such that the mean drag coefficient also increases. Nevertheless, other mechanisms cannot be ruled out, since, for example, Faraday waves have been shown to greatly enhance acoustic streaming~\cite{Maksimov07}. Hence, new models and numerical simulations are necessary to the full understanding of the observed rise velocity reduction.


To conclude, our observations clearly establish the strong impact of surface Faraday waves on a bubble's buoyant rise. Such waves are most easily triggered by resonant acoustic forcings such that moderate sound levels can induce important rise velocity modifications. This suggests to take into account the acoustic noise in multiphase flow problematics: air entrainment in oceanic surf zones~\cite{Dean97}, charge losses in bubbly flows, or turbulence drag reduction by microbubble injection~\cite{Lo2006}. From a more applied point of view, our observations 
could initiate new strategies for industrial processes where the bubble rise rate is crucial, or for soft-matter sorting by acoustophoresis.
Indeed, as shown in a recent paper~\cite{Cross07}, some cancer cells have a lower membrane elasticity and this contrast in stiffness should modify surface waves triggered on the cell wall, which in turn modifies the flow properties of the cell. It is the subject of ongoing investigations.

The authors wish to thank E. Hervieu and M. Guinard for scientific support and stimulating discussions as well as C. Kevorkian for helpful technical support.

 \end{document}